\documentclass[12pt,a4paper]{amsart} %jesli chce numeracje jednostronnq, to daje ,oneside, w graniaste.
\usepackage{amsmath}
\usepackage{amsthm}
\usepackage{amssymb}
\usepackage{ot1patch}
\usepackage{fancyhdr}

\usepackage{graphicx}

\usepackage{float}
\restylefloat{table}

\theoremstyle{remark}

\newtheorem*{rem*}{Remark}

\theoremstyle{definition}

\numberwithin{equation}{section}

\begin{document}
\title{Linkages and systemic risk in the European insurance sector:
Some new evidence based on dynamic spanning trees}

\author{Anna Denkowska}\address{Cracow University of Economics, Chair of Mathematics, Rakowicka 27, 31-510 Krak\'ow, Poland}\email{anna.denkowska@uek.krakow.pl}\author{Stanis\l aw Wanat}\address{Cracow University of Economics, Chair of Mathematics, Rakowicka 27, 31-510 Krak\'ow, Poland}\email{stanislaw.wanat@uek.krakow.pl}

\begin{abstract}
This paper is part of the research on the interlinkages between insurers and their contribution to systemic risk on the insurance market. Its main purpose is to present the results of the analysis of linkage dynamics and systemic risk in the European insurance sector which are obtained using correlation networks. These networks are based on dynamic dependence structures modelled using a copula. Then, we determine minimum spanning trees (MST). Finally, the linkage dynamics is described by means of selected topological network measures. 
\end{abstract}

\maketitle

\section{Introduction}
This article is devoted to the dynamics of the interconnectedness of some among the largest European insurers. Our main tool are the minimum spanning trees. The results obtained are presented in the context of systemic risk on the European insurance market. It should be pointed out that after the financial crisis during the years 2007-2009 and the European public debt crisis during the period 2010-2012, both the academic community and the supervisory authorities have started to pay more attention to the role played by insurance institutions in creating systemic risk. Before that, researchers seemed to be convinced that the insurance market is systemically irrelevant. After the crisis, some of them upheld their point of view:  \cite{Harrington}, \cite{Bell Keller},  \cite{Geneva}, \cite{Bednarczyk}, while others published papers indicating that systemic risk may be created by the insurance sector: \cite{Billio}, \cite{WM},\cite{Baluch}, \cite{Cummins Weiss}, \cite{Chen}, \cite{Czerwinska}. In \cite{Bierth} basing on the study of a large number of insurers in a long timescale, the authors come to the conclusion that the contribution of the insurance sector to systemic risk is relatively small and its peak was reached during the financial crisis of the years 2007-2008. They also indicate the four L's: linkages between large insurance companies, leverage, losses, liquidity, as important factors for the insurers' exposure to systemic risk.

After the crisis, the supervisory authorities, too, reached the conclusion that systemic risk may be generated on the insurance market. As a result they developed a method allowing to distinguish the insurance institutions that have a particular impact on financial stability\footnote{This list is published by FSB (Financial Stability Board) (see {\tt https://www.fsb.org/work-of-the-fsb/policy-development/addressing-sifis/} {\tt global-systemically-important-financial-institutions-g-sifis/}). Currently it consists of: Aegon, Allianz, AIG, Aviva, AXA, MetLife, Ping An Insurance (Group) Company of China, Prudential Financial Inc., Prudential plc.}. This method takes into account the following five dimensions cf. \cite{IAIS}\footnote{The brackets contain the weight associated to each  quantity when computing in the general index.}: \begin{itemize}
\item the size of the insurance institution (5\%),
\item its range of activities of global character (5\%),
\item assessment of the degree of direct and indirect linkages between institutions in the financial system (40\%),
\item non-traditional activities of the insurer outside the insurance sector (45\%),
\item product substitutability  -– the institution's significance grows together with the lack of real possibilities of substitution for the services rendered by the insurer (5\%).
\end{itemize}
Therefore, we can say that both theoreticians and practitioners are convinced that insurance institutions have a potential of creating systemic risk and that the interconnectedness of insurance companies is one of the major factors influencing risk exposure.

The report \cite{EIOP} indicates that in order to assess potential systemic risk one needs to consider the build-up of risks, including the risks that are built up over time, as well as the interconnectedness within the financial sector and the wider economy. In the same report it was remarked that linkages in the insurance sector and between the latter and other parts of the financial sector, most notably the banking one, need to be addressed and analyzed. Our present article is an attempt to respond to these challenges. We concentrate on the study of the dynamics of the interconnectedness structure of the insurers in the propagation of systemic risk on the insurance market. This dynamics has been studied using properly chosen topological indices of minimum spanning trees built using conditional correlations of returns of European insurers. Indeed, we believe that a higher correlation of insurers' stock prices implies that more insurers are exposed to the same kind of turmoil at the same time and they will tend to react more similarly when hit by a shock.

The contribution of our article is two-fold. Firstly, it is the application of two-dimensional copula-DCC-GARCH models to estimate the conditional correlation coefficients, crucial for the analysis. Secondly, we present the result of the analysis of the interconnectedness structure dynamics in the European insurance sector obtained using selected topological indicators for minimum spanning trees and we check, if during high turbulence periods on the financial market the network's structure is different from the structure observed during the periods when the market was in its `normal state'. To the best of our knowledge, this approach is novel and has not been used in the literature yet.

The paper is organized as follows. In the second chapter we discuss briefly the topological indicators or indices we shall be using. In the third --- we present the empirical strategy, while the fourth one is devoted to the data and the results obtained. Finally, we draw our conclusions in the fifth and last chapter.

\section{Methodology --- topological indices of networks}

One of the newest mainstreams of research in interconnectedness on the insurance market makes use of graph theory. The Minimum Spanning Tree, which we will abbreviate MST, is a major tool taken from this theory. The MSTs are widely used due to their good filtering and compression properties in the case of complex systems having a network structure, which simplifies the description and analysis of the processes that take place, see \cite{Mantegna}, \cite{MantegnaS}. At the same time, in order to study the structure of the network of linkages when this structure evolves in time, researchers avail themselves of time series of adequate topological indices of the network found thanks to the MST constructed for each period studied. In practice, the most used time series involve the following indices:\begin{itemize}
\item The Average Path Length -- APL. This index is defined as the average number of steps taken along all the shortest paths connecting all possible pairs of network nodes. It measures the effectivity of information flow or mass transport of a network. The APL is one of the most robust measures of network topology (alongside with the clustering coefficient and the degree distribution). It is clear that APL will tell an easily negotiable network from an inefficient or complex one. However, although the smaller the APL, the better the diffusion of information, we should keep in mind that as we use an average quantity, we can get a small APL also for a network that has several very distant nodes and many neighbouring ones. 

If we consider a network as an unweighted directed graph with a set of $n>1$ vertices $V$ and put $\delta(u,v)$ for the length of the shortest path connecting the two distinct vertices $u, v$ (with the convention that it reduces to zero in case they cannot be connected), then the APL is given by the formula
$$
\frac{1}{n(n-1)}\sum_{u,v\in V\colon u\neq v}\delta(u,v).
$$
\item The Maximum Degree: in graph theory it is defined as the maximal number of edges coming out from a vertex (where each loop counts for two). In other words, it measures the number of connections to the central vertex. 
\item The Betweenness Centrality -- abbreviated BC. It measures the {\it centrality} of a vertex: we consider the ratio between the number of shortest paths connecting two vertices and passing through the given one, and the number of all the shortest paths between pairs of distinct vertices. It indicates thus the most important nodes of a network based on shortest paths (e.g. the most influential insurer). For each pair of vertices of a connected graph there always exists at least one path connecting them and such that either the number of edges it passes through (for unweighted graphs), or the sum of the weights (for weighted graphs) is minimized.  In plain words, BC gives the number of the shortest paths passing through a fixed vertex and therefore it specifies to what extent a given node serves as an intermediary for other nodes of the network. In particular, a node with high BC has more control over the network.
\item The parameter $\alpha$ from the degree distribution required to follow (asymptotically) a power law -- i.e. it concerns scale-free networks. This parameter measures the scale-free behaviour of the network. To be more precise, if we denote by $P(k)=\frac{n_k}{n}$ the degree distribution where $n_k$ is the number of  vertices of the graph having degree $k$ and $n$ the total number of vertices, then we require that $P(k)\sim Ck^{-\alpha}$, where $\alpha>0$ is a parameter specific to the given network. The power law followed by $P(k)$ results in the network having some (fractal) self-similarity properties which accounts for the name {\it scale-free}. Such networks have typically a small number of nodes having many connections (such nodes are called {\it hubs}) and a large number of nodes with only a single connection. From the point of view of our analysis this kind of network is considered as favorable to the diffusion of information (systemic risk) and the hubs it contains are systemically relevant. 
\end{itemize}
We should add that in the literature MSTs that evolve in time are monitored also through many other topological indices such as e.g. the normalized tree length, the mean occupation layer, the tree half-life \cite{Onnela2003};  survival ratio of the edges \cite{Onnela2002}, \cite{Sensoy}; node degree, strength \cite{Sensoy}; eigenvector \cite{Tang}; closeness centrality \cite{Sensoy} and agglomerative coefficient \cite{MO}.
  
\section{Empiric strategy}

The empiric strategy we use in the present paper consists essentially of two stages:
\begin{itemize}
\item first we construct the minimum spanning tree $MST_t$ for each period $t=1,\dots,T$ under consideration;
\item then, using $MST_t$ we define the time series of the chosen topological indices of the network.
\end{itemize}
The first stage is based on the usual procedure well-known from the literature (cf. \cite{Marti} and the references therein):
\begin{enumerate}
\item We determine the logarithmic return rates $r_{i,t}$, $i=1,\dots,k$, $t=1,\dots, T$, based on the stock quotes of $k$ insurers.

\item Using the logarithmic return rates $r_{i,t}$ we estimate the conditional linear correlation coefficients $R_t(i,j)$ for each pair of insurers $(i,j)$ $i,j=1,\dots, k$ and each period $t=1,\dots, T$. 

\item For each period we determine the matrix of distances between the insurers, making use of the following metric from \cite{Mantegna}: 
$$
d_t(i,j)=\sqrt{2(1-R_t(i,j))}
$$
where $R_t(i,j)$ is the correlation coefficient between the $i$-th and $j$-th insurers, $i,j=1,\dots, k$, for $t=1,\dots, T$.

\item Next, using the distance matrices we construct the minimum spanning trees $MST_t$ ($t=1,\dots, T$) with $k$ vertices and $k-1$ edges, thanks to the Kruskal algorithm, cf. \cite{MantegnaS}. The obtained graph $MST_t$ is a model of the network of connections between insurers for the period $t$. Its vertices represent the different insurers, whereas the edges connect those pairs of them that share the most of similarities (connection with an edge means that the distance, as defined above, between the two vertices is relatively small). In some sense, this construction amounts to finding the most convenient $k-1$ connections among all the $k(k-1)/2$ connections available. In the case that we analyze, the $MST_t$'s can be seen as filtered networks that allow the identification of the most probable and the shortest path of crisis transmission (of systemic risk).
\end{enumerate}

The correlation coefficients $R_t(i,j)$, crucial for that procedure, are obtained using two-dimensional copula-DCC-GARCH models estimated for each pair of insurers. To the authors' best knowledge, this approach has never been applied to construct dynamic minimum spanning trees before.

In the copula-DCC-GARCH model, the distribution of the vector $r_t=(r_{1,t},\dots, r_{k,t})$ of return rates, conditional w.r.t. the set $\Omega_{t-1}$ of information available up to the time period $t-1$, is modeled using the conditional copul{\ae} proposed by Patton in \cite{Patton}. It takes the following form:
\begin{align*}
&r_{1,t}|\Omega_{t-1}\sim F_{1,t}(\cdot\mid \Omega_{t-1}),\ldots, r_{k,t}|\Omega_{t-1}\sim F_{k,t}(\cdot\mid\Omega_{t-1})\\
&r_t|\Omega_{t-1}\sim F_t(\cdot\mid\Omega_{t-1})\\
&F_t(r_t\mid \Omega_{t-1})=C_t(F_{1,t}(r_{1,t}\mid \Omega_{t-1}),\ldots, F_{k,t}(r_{k,t}\mid\Omega_{t-1}))
\end{align*}
where $C_t$ stands for the copula, while $F_t$ and $F_{i,t}$ denote the distribution function of the multivariate distribution and the the distribution function of the marginal distributions at time $t$, respectively. In the general case, one-dimensional return rates can be modeled using various specifications of the mean model (e.g. the ARIMA process) as well as various specifications of the variance model (e.g. sGARCH, fGARCH, eGARCH, gjrGARCH, apARCH, iGARCH, csGARCH). In our study, we used the following ARIMA process for all the average return rates series:
\begin{align*}
&r_{i,t}=\mu_{i,t}+y_{i,t},\\
&\mu_{i,t}=E(r_{i,t}\mid\Omega_{t-1}),\\
&\mu_{i,t}=\mu_{i,0}+\sum_{j=1}^{p_i}\varphi_{ij}r_{i, t-j}+\sum_{j=1}^{q_i}\theta_{ij}y_{i,t-j},\\
&y_{i,t}=\sqrt{h_{i,t}}\varepsilon_{i,t},
\end{align*}
whereas for the variance we applied the exponential GARCH (eGARCH) \cite{Nelson}:
$$
\log(h_{i,t})=\omega_i+\sum_{j=1}^{p_i}(\alpha_{ij}\varepsilon_{i,t-j}+\gamma_{ij}(|\varepsilon_{i,t-j}|-E|\varepsilon_{i,t-j}|))+\sum_{j=1}^{q_i}\beta_{ij}\log(h_{i,t-j})
$$
where $\varepsilon_{i,t}=\frac{y_{i,t}}{\sqrt{h_{i,t}}}$ are identically distributed independent random variables (in the empirical study we consider the normal distribution, the skew normal distribution, t-Student, skew t-Student and GED).

In order to describe the dependances between reutrn rates we use t-Student copul{\ae} with the conditional correlations $R_t$ obtained from the model $DCC(m,n)$ as parameters:
\begin{align*}
&H_t=D_tR_tD_t,\\
&D_t=\mathrm{diag}(\sqrt{h_{1,t}},\dots,\sqrt{h_{k,t}}),\\
&R_t=\left(\mathrm{diag}(Q_t)\right)^{-1/2}Q_t\left(\mathrm{diag}(Q_t)\right)^{-1/2},\\
&Q_t=\left(1-\sum_{j=1}^mc_j-\sum_{j=1}^nd_j\right)\bar{Q}+\sum_{j=1}^mc_j(\varepsilon_{t-j}\varepsilon'_{t-j})+\sum_{j=1}^nd_jQ_{t-j}.
\end{align*}
Here $\bar{Q}$ is the unconditional covariance matrix of the standardized residuals $\varepsilon_t$, $c_j$ ($j=1,\dots, m$) are scalars describing the influence of precedent shocks on the current correlations while the scalars $d_j$ ($j=1,\dots,n$) represent the influence of the precedent conditional correlations.

In the second stage we determine, using the minimum spanning trees $MST_t$ obtained, the time series for the following topological network indices:\begin{itemize}
\item the average path length APL,
\item the maximum degree,
\item the parameters $\alpha$ of the power law of the degree distribution,
\item the betweenness centrality BC.
\end{itemize}

\section{Data and analysis results}

As a a basis for our study we took the stock quotes of 28 European insurance institutions chosen among 50 largest such institutions\footnote{From the 50 largest we chose the companies that were listed during the period studied i.e. 07.01.2005-26.04.2019.} according to {\tt https://www.relbanks.com/top-insurance-companies/europe} (see Table 1). We analyzed weekly logarithmic return rates for the time period from January 7th 2005 to April 26th 2019.
{\small
\begin{table}[H]\caption{\small Insurance companies considered in the study and their abbreviations used in the presentation of the results. (Source: authors' own elaboration based on {\tt https://www.relbanks.com/top-insurance-companies/europe}.)}\centering
\begin{tabular}{|r|c|c|c| p{2cm}|p{1cm}|}
\hline
No.&Insurer &Abbreviation &Country &Total assets in bil. USD (balance sheet 12/31/2016) &Place in the ranking\\ \hline
1 & AXA & AXA & France & 944,145 &1\\ \hline
2 &Allianz & Alli &Germany & 934,654 &2\\ \hline
3 &Prudential plc &Prud &Great Britain & 578,149 & 3\\ \hline
4 &Legal \& General &Lega & Great Britain & 574,901 & 4\\ \hline
5 &Generali &Gene & Italy &551,168 &5\\ \hline
6 &Aviva &Aviv &Great Britain & 541,188 &6\\ \hline
7 & Aegon &Aego & Netherlands &450,439 &7\\ \hline
8 &CNP Assurances &CNP &France &443,242 &8\\ \hline
9 &Zurich Insurance &Zuri &Switzerland &382,679 &9\\ \hline
10 &Munich Re &Mu.Re &Germany & 283,206 &10\\ \hline
11 &Old Mutual &Ol.Mu & Great Britain &210,823 &13\\ \hline
12 &Swiss Life &Swiss &Switzerland &196,373 &14\\ \hline
13 & Chubb Ltd &Chub & Switzerland & 159,786 &17\\ \hline
14 & Ageas &Agea &Belgium & 110,294& 19\\ \hline
15 &Phoenix &Phoen & Great Britain &105,676 &20\\ \hline
16 &Unipol Gruppo &Unip &Italy &97,184& 23\\ \hline
17& Mapfre & Mapf& Spain& 71,787& 26\\ \hline
18& Hannover Re&Hann&Germany&67,184& 28\\ \hline
19& Storebrand &Stor &Norway& 60,508& 29\\ \hline
20& XL.Group& XL.Gr &Bermuda& 58,434& 30\\ \hline
21 &Helvetia Holding &Helv & Switzerland &54,299 &31\\ \hline
22 &Vienna Insurance &Vien& Austria & 52,981 &32\\ \hline
23 &SCOR SE &SCOR &France &45,784 &33\\ \hline
24 &Mediolanum &Medi & Italy& 44,386 &34\\ \hline
25 &Sampo Oyj &Samp &Finland &40,139& 35\\ \hline
26& RSA Insurance Group& RSA & Great Britain& 25,976& 39\\ \hline
27& Societ\`a Cattolica di Assicurazione&So.Ca &Italy&25,627&40\\ \hline
28& Topdanmark A/S &Topd& Denmark &10,451&47\\ \hline
\end{tabular}
\end{table}
}
Proceeding to construct the minimum spanning tree, we computed the correlation coefficients $R_t(i,j)$ from the two-dimensional copula-DCC-GARCH models estimates for each pair of insurers. During this analysis we used different specifications ARMA-GARCH of one-dimensional models. Eventually, basing on information criteria and adequacy tests, we chose for all the returns the ARMA(1,1)-eGARCH(2,2) model with the skew t-Student distribution. In the analysis of the dependances' dynamics we considered the Gauss and Student Copul{\ae} togehter with various specifications of the DCC model. As earlier, from information criteria we chose the Student copula with correlation conditionals obtained from the DCC(1,1) model and a constant shape parameter (estimation results for all the 372 models available on demand). On the basis of the correlation coefficients obtained in this way for each period $t=1,\dots, 747$ studied we determined the distance matrices. Then, using the Kruskal algorithm, we constructed the minimum spanning tree. Four such trees for the beginning and the end of two periods when the network was shrinking (clear drop in the APL), i.e. 02.06.2006--17.08.2007 and 05.12.2008--17.09.2010, can be seen on Figure 1. 
\begin{figure}[H]
\centering
\includegraphics[width=11 cm]{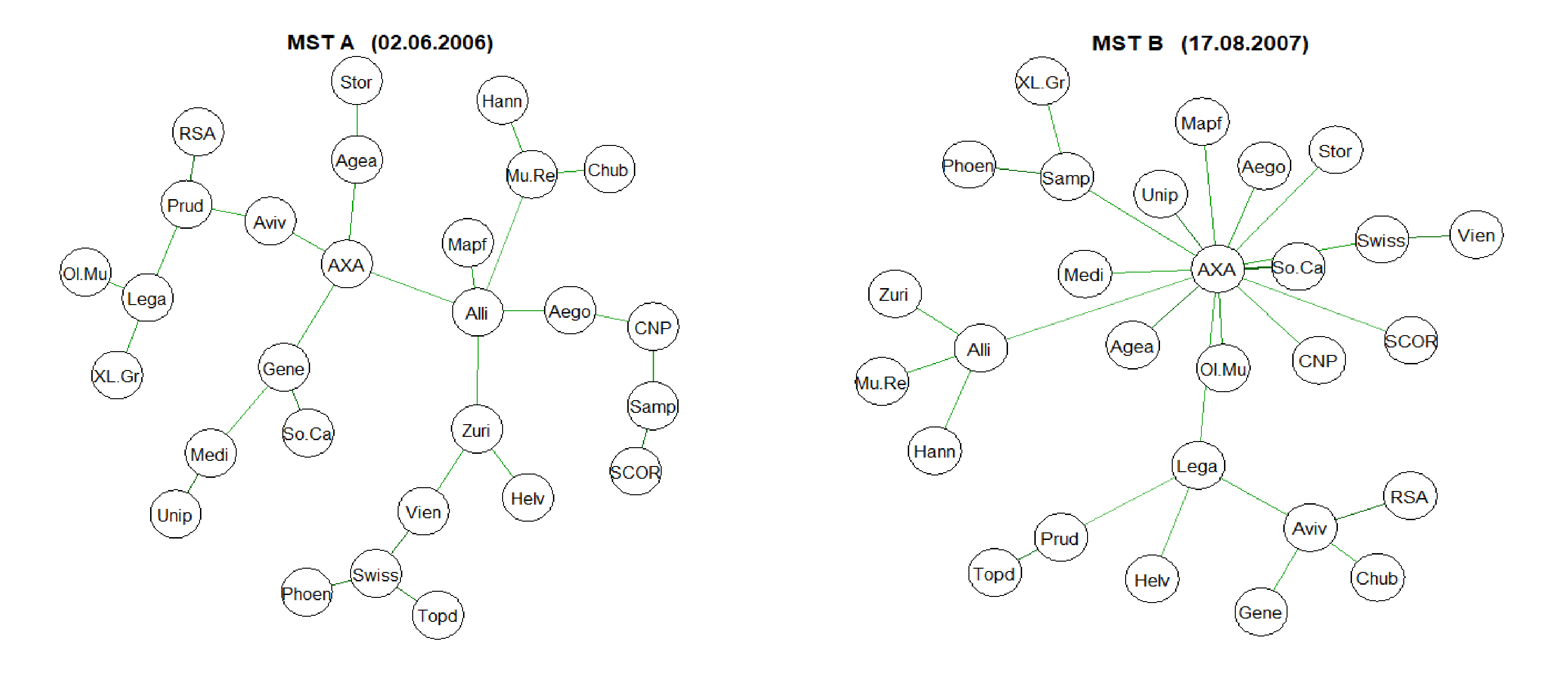}
\includegraphics[width=11 cm]{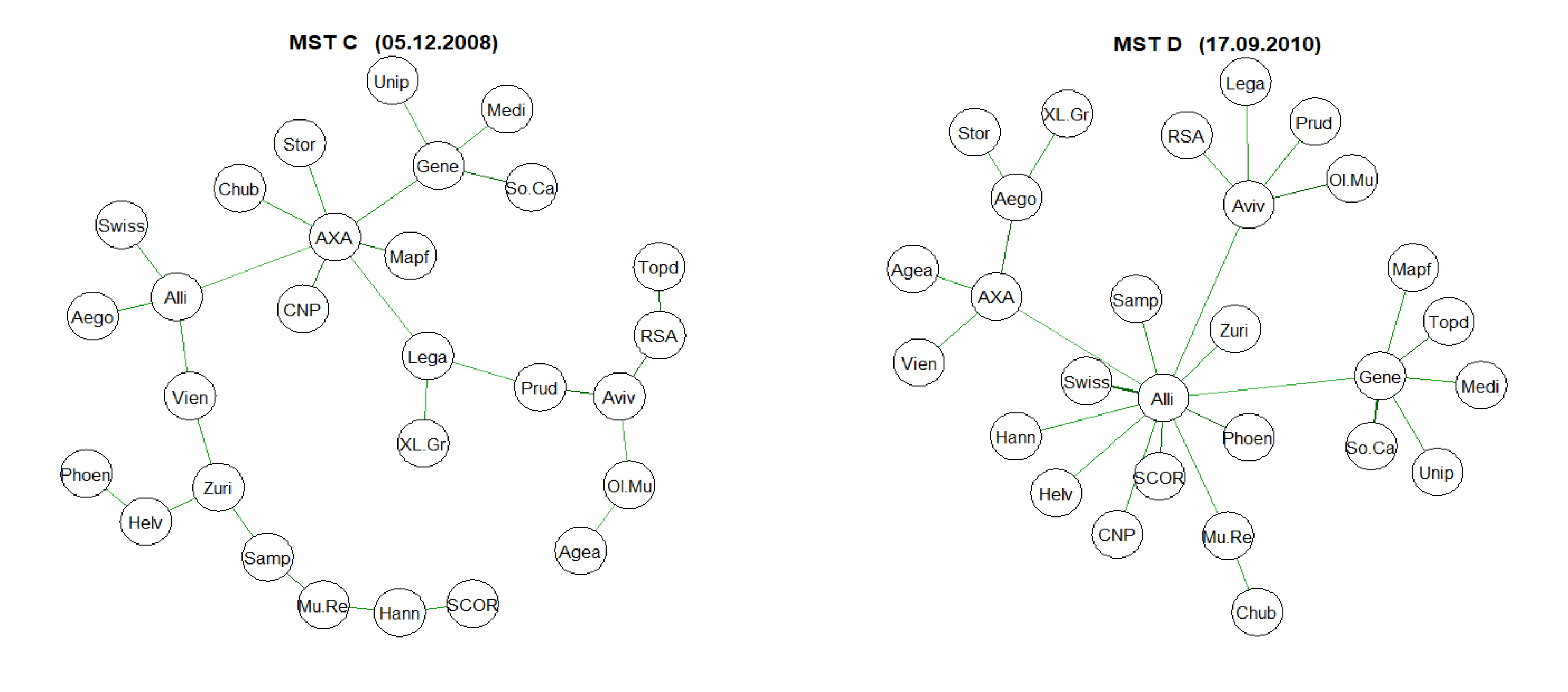}
\caption{\small Minimum spanning trees in the beginning and at the end of two periods when the network was shrinking, i.e. 02.06.2006-17.08.2007 and 05.12.2008-17.09.2010. (Source: authors' own elaboration.)}
\end{figure}

In the second stage the miminum spanning trees obtained were used to determine the time series for: the average path legth (Fig. 2), the maximum degree (Fig. 3), the parameters $\alpha$ of the power law of the degree distribution and the corresponding values of pValue (Fig. 4) and the betweeness centrality. The average values of BC for the studied period 07.01.2005--26.04.2019 obtained for the different insurers is shown on Fig. 5. Finally, Figure 6 presents the BC times series only for selected insurers, i.e. for the two having the highest, lowest and `average' mean values of the indicator. The highest values $0,709$ and $0,549$ were obtained for AXA and Allianz, respectively, the lowest (0,000) -- for Phoenix, Chubb Ltd and XL.Group, and the `average' mean value of BC is represented by Munich Re.

\begin{figure}[H]
\centering
\includegraphics[width=11 cm]{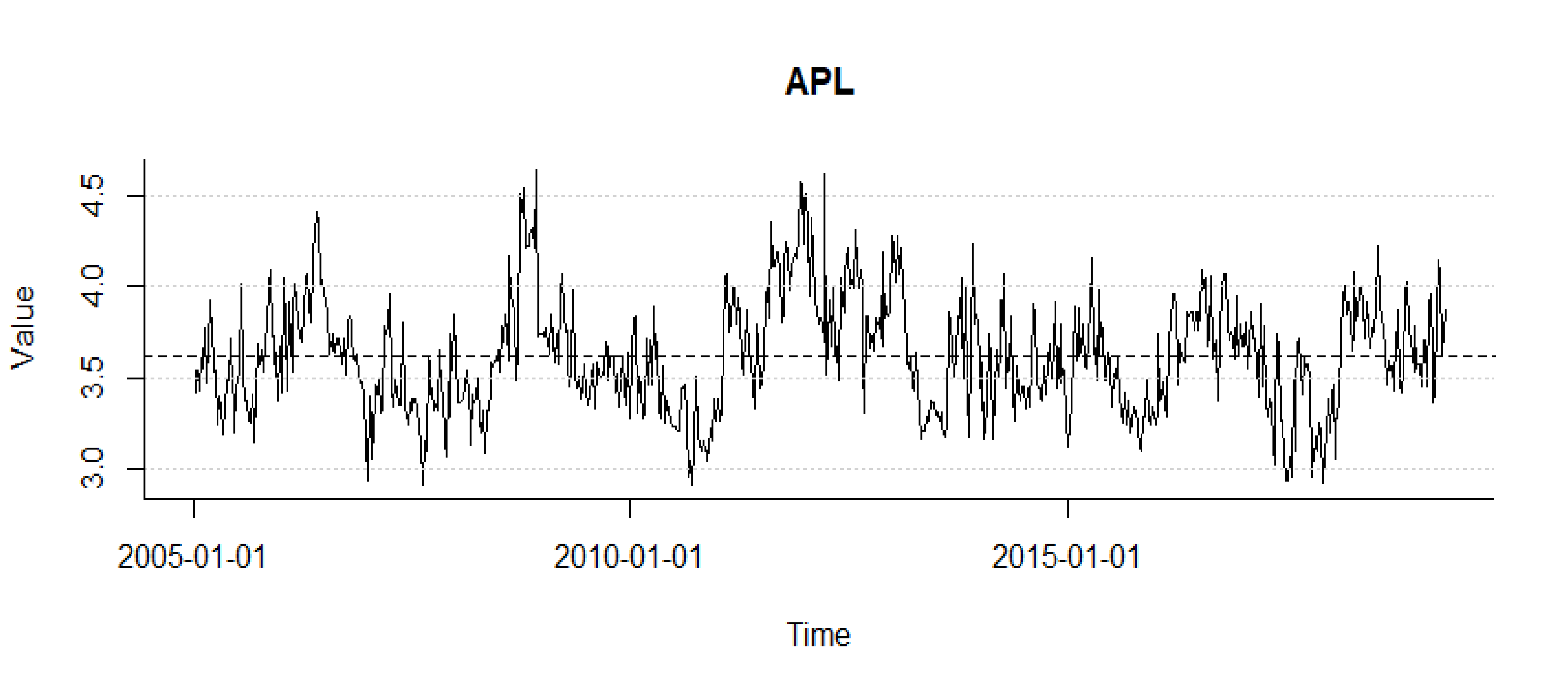}
\caption{\small Average path length for minimum spanning trees during the period studied (07.01.2005-26.04.2019). (Source: authors' own elaboration.)}
\end{figure}

\begin{figure}[H]
\centering
\includegraphics[width=11 cm]{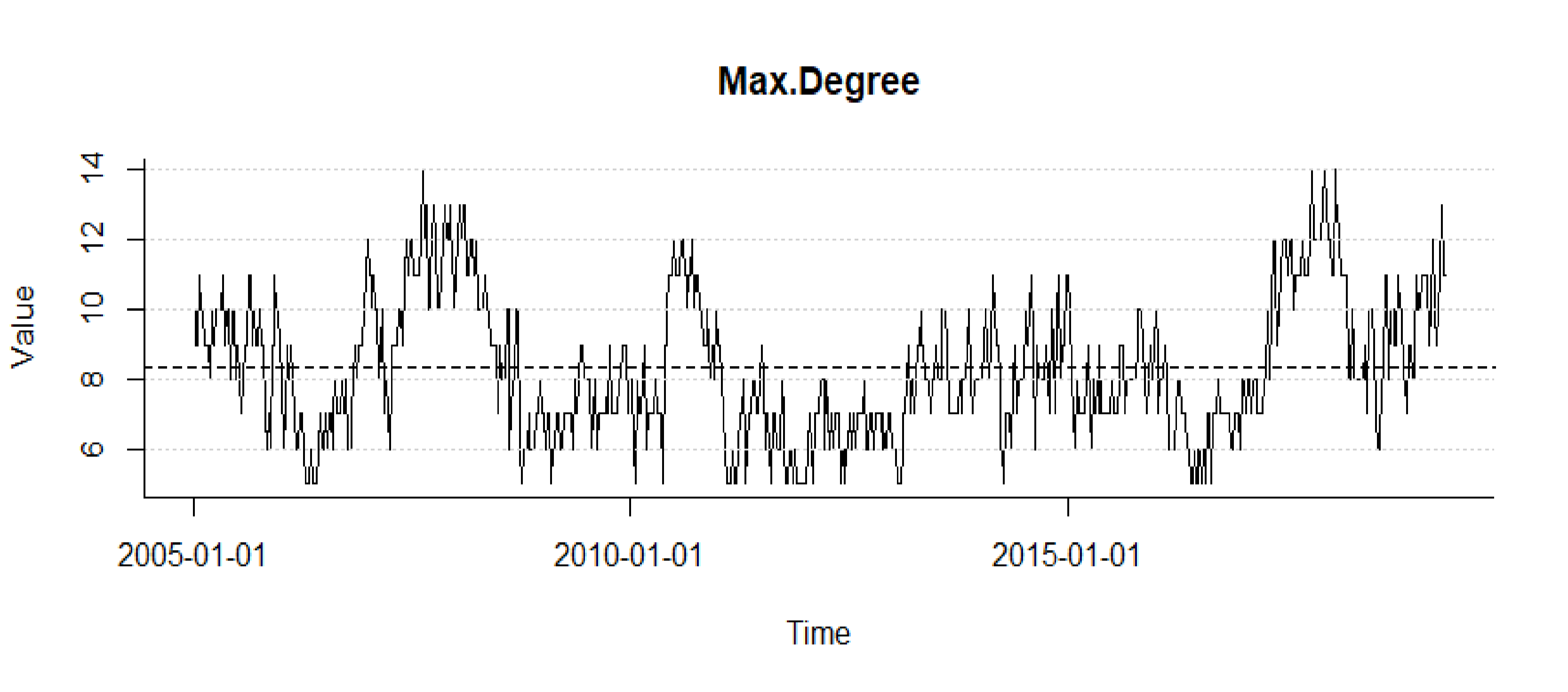}
\caption{\small Maximum degrees for minimum spanning trees during the period studied (07.01.2005-26.04.2019). (Source: authors' own elaboration.)}
\end{figure}

\begin{figure}[H]
\centering
\includegraphics[width=11 cm]{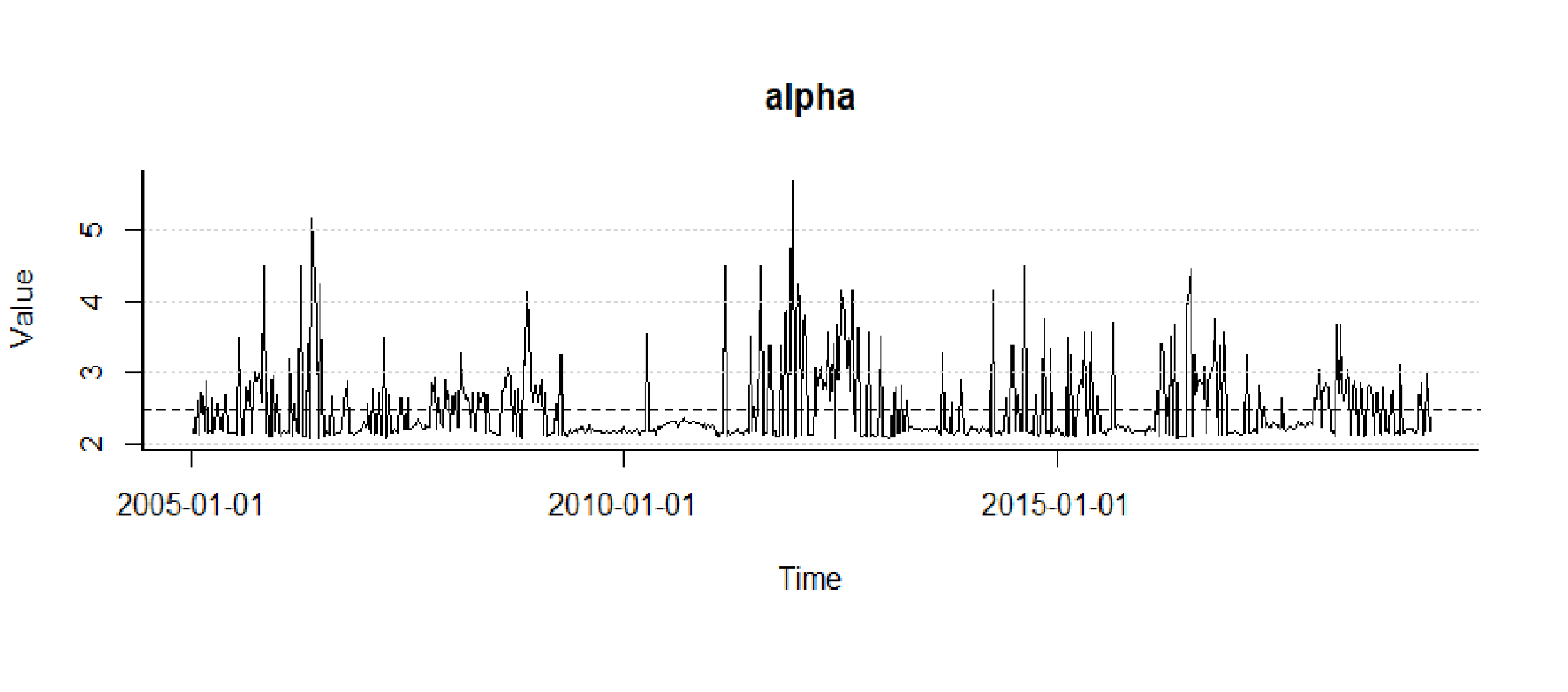}
\includegraphics[width=11 cm]{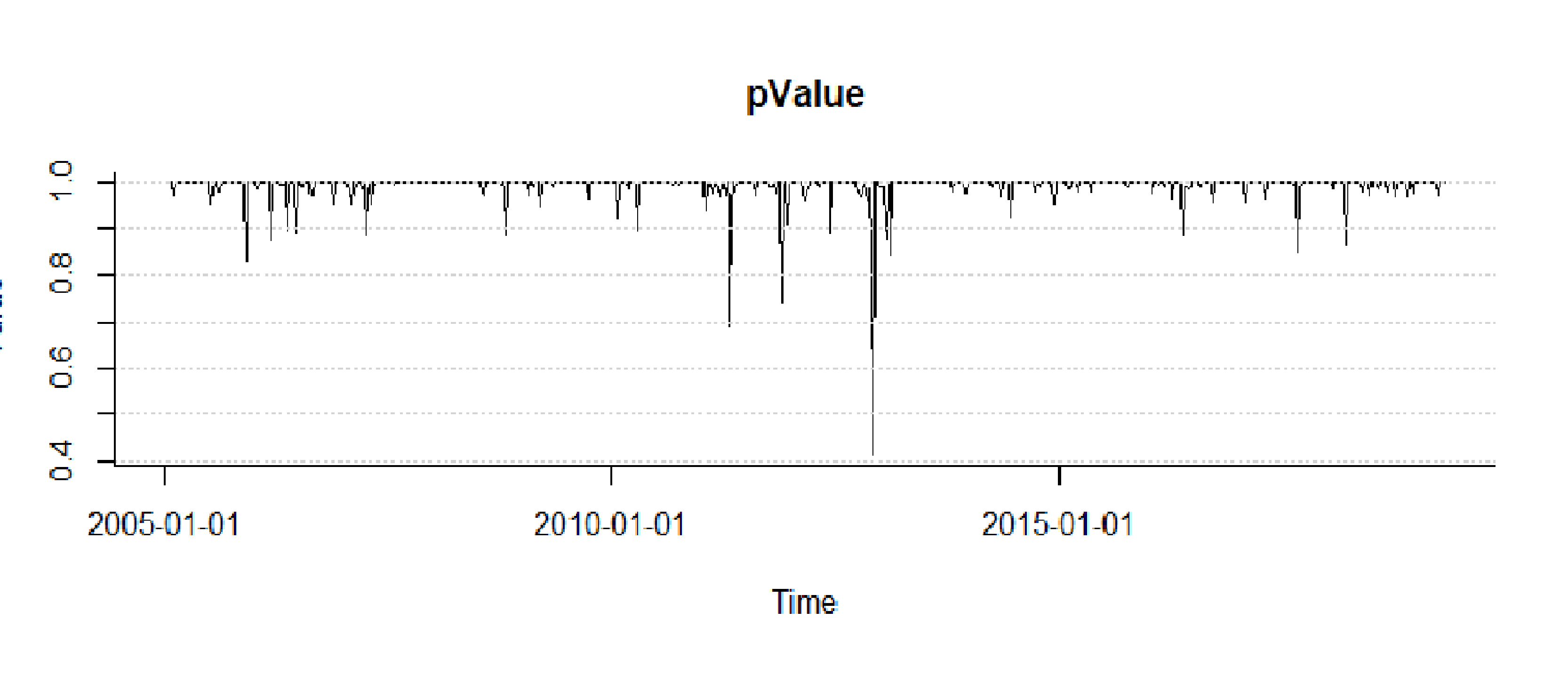}
\caption{\small Estimated parameters $\alpha$ of the power law and the corresponding values pValue for the MST during the period studied (07.01.2005-26.04.2019). (Source: authors' own elaboration.)}
\end{figure}

\begin{figure}[H]
\centering
\includegraphics[width=12 cm]{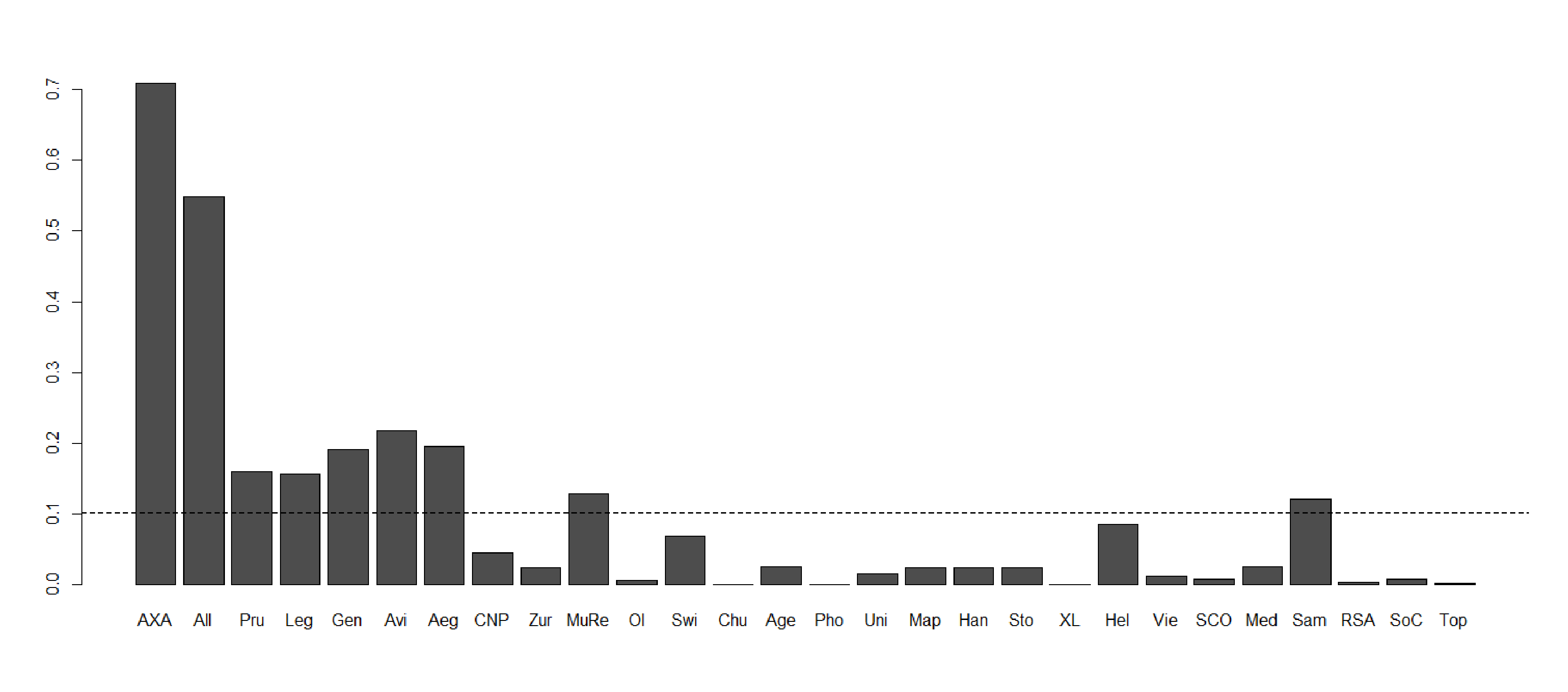}
\caption{\small Mean value of BC during the period studied 07.01.2005-26.04.2019 for each insurer. (Source: authors' own elaboration.)}
\end{figure}

\begin{figure}[H]
\centering
\includegraphics[width=11 cm]{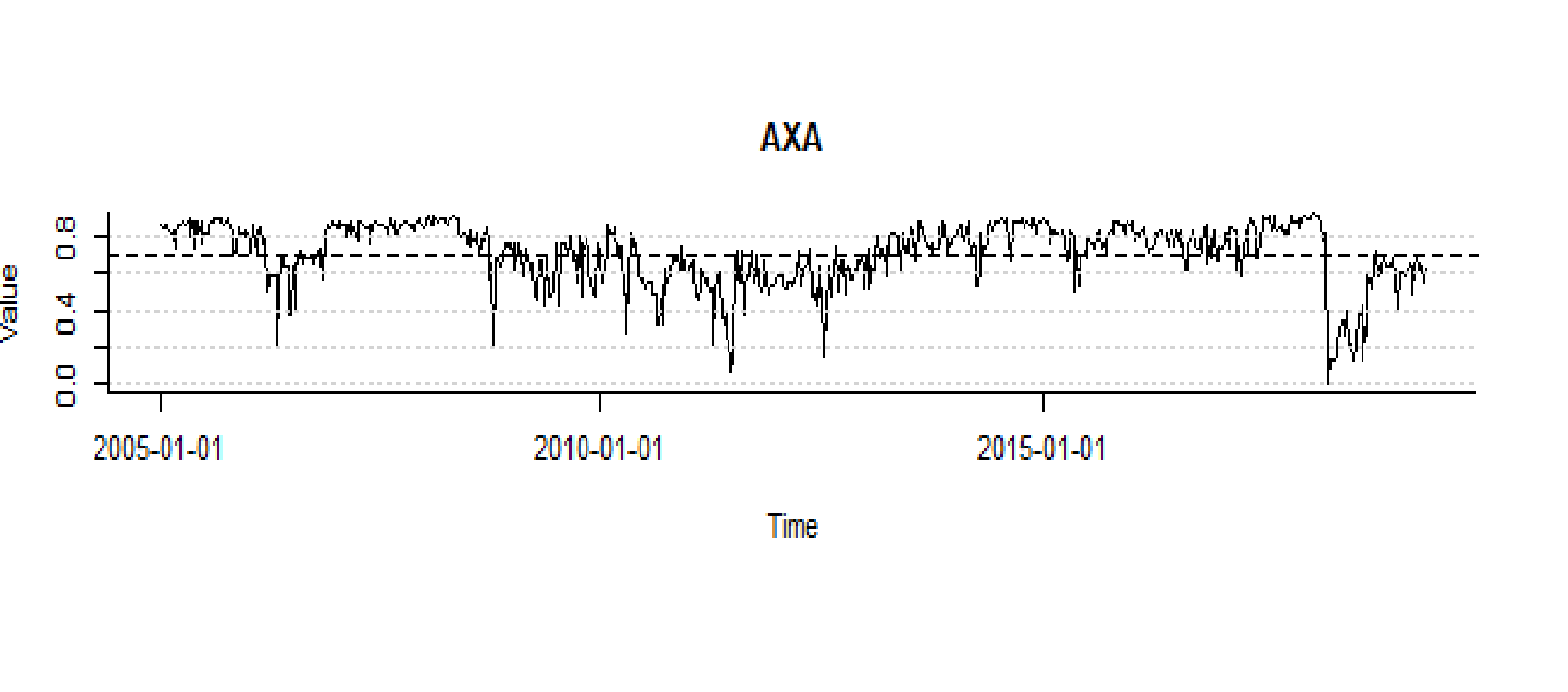}
\includegraphics[width=11 cm]{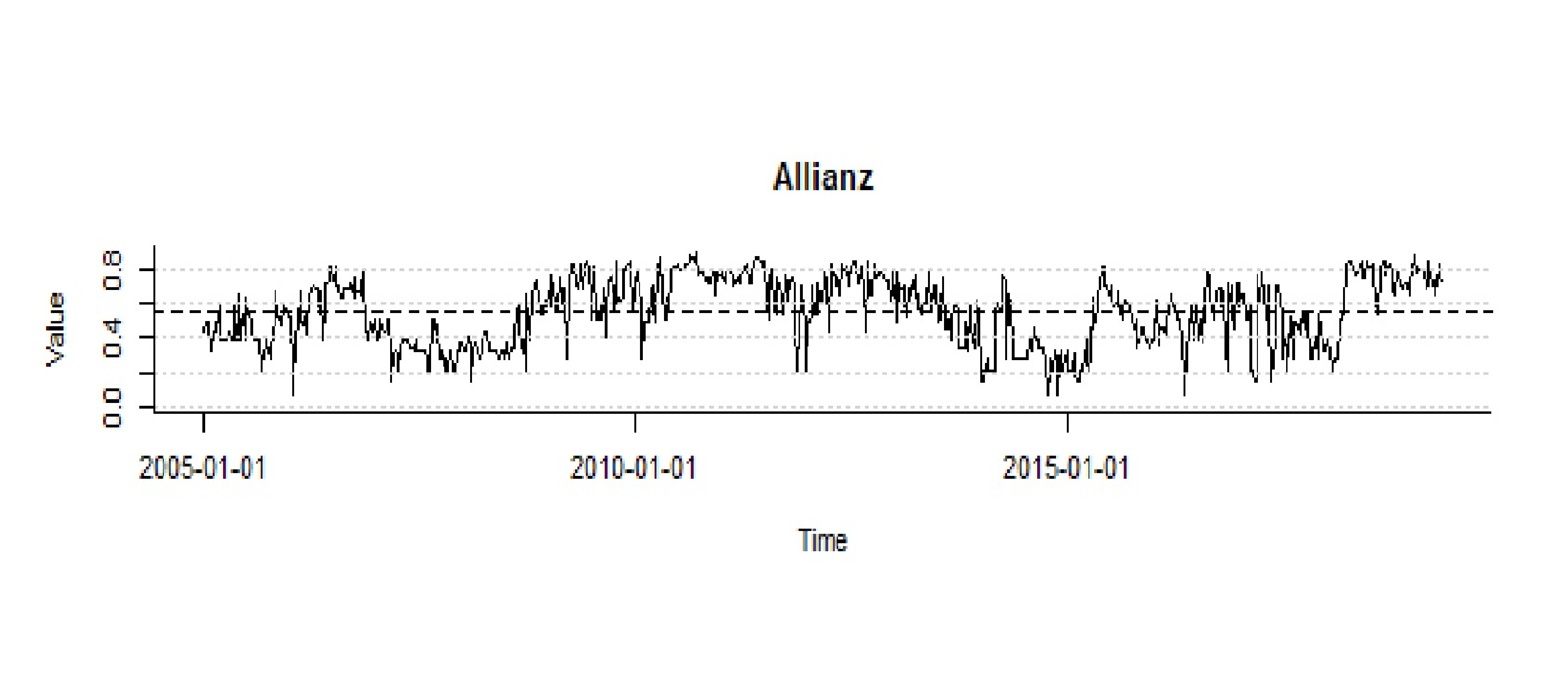}
\includegraphics[width=11 cm]{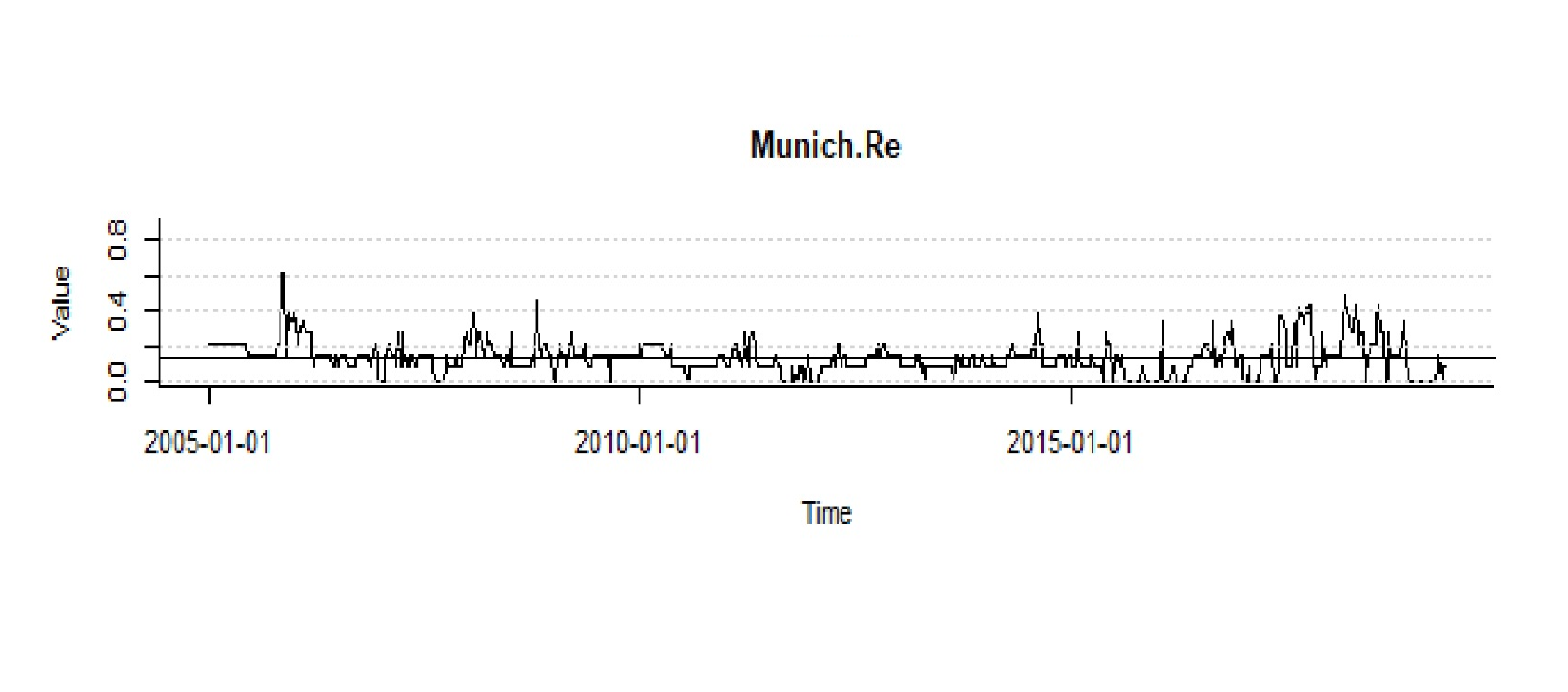}
\includegraphics[width=11 cm]{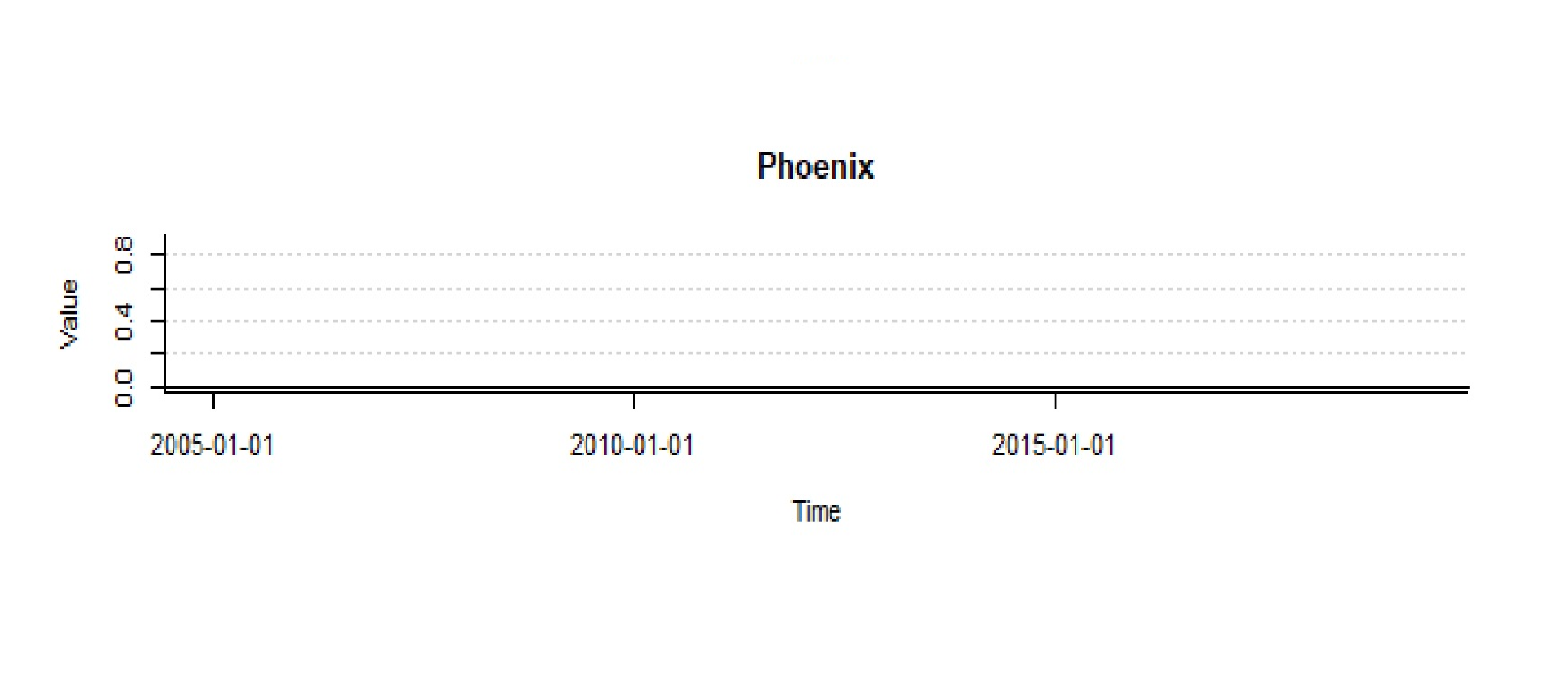}
\caption{\small BC of selected insurance companies (i.e. AXA, Munich Re and Phoenix) during the period studied (07.01.2005-26.04.2019). (Source: authors' own elaboration.)}
\end{figure}

\section{Conclusions}

Our aim was to study the dependance and linkage structure, together with its dynamics during the period 07.01.2005--26.04.2019 enclosing the financial crisis of the years 2007--2009 as well as the European public debt crisis of 2010--2012, in the case of the largest European insurance companies and making use of the minimum spanning trees. The correlation coefficients, crucial for the analysis made, were obtained using the estimated two-dimensional copula-DCC-GARCH models, whilst the minimum spanning trees were constructed by means of the Kruskal algorithm. The dependance structure dynamics was studied with the aid of time series for suitably chosen topological indices of the network. 

The results lead to the conclusion that the analyzed time series of the topological indices for the minimum spanning trees $MST_t$ ($t=1,\dots,T$) constructed using the distances obtained from the dynamical correlations, do not show any trend, but at the same time they show a significant variability.

Looking on the average path length and the maximum degree (cf. Figures 2 and 3) we can determine the periods during which these two indicators are clearly below the average (shown as the dotted line on the two figures), and the periods during which their values are distinctly above average. Moreover, it is apparent that a drop in the APL goes together with a soar of the maximum degree. When the mean distance tends to get smaller, it signalizes a shrinking of the dependances network, i.e. an uplift of the interdependances between the insurance companies which implies also a higher ability of information transfer. A clear shrinking of the network can be seen for the period 02.06.2006--17.08.2007, i.e. precisely just before the subprime crisis and during its first phase\footnote{The crisis phases and the developmentsl inked to them are described in details in \cite{Geneva}, Appendix A. Timeline of crisis.}, as well as for the period 05.12.2008--17.09.2010, i.e. just before and at the very beginnning of the European public debt crisis. The minimum spanning trees at the beginning and at the end of these two periods are shown on Fig. 1 (MST A, MST B,  MST C, MST D, respectively). On that basis, analyzing the minimum spanning trees structure for the remaining weeks (the corresponding results are available on demand) we can come to the conclusion that during the time of the largest turbulences on global markets, the average distance for the trees is low, while their degree is high. In this kind of trees we can distinguish several crucial insurance companies with a high value of betweenness centrality, i.e. having an important control over the network. As it is apparent from the graphics shown on Fig. 6, during the subprime crisis, the network was controlled essentially by AXA alone, whereas during the European public debt crisis, Allianz took over the control. In general, we can also remark that during the period studied, these are the two companies that gain in turn control over the network.

On the other hand, from the analysis of the time series for the parameters $\alpha$ and the corresponding values pValue we can see that in each period studied the degree distribution of $MST_t$ follows a power law (see Fig. 4), i.e. the networks are scale-free. Typically, such networks have a small number of vertices with a large number of edges starting from them --- these are the hubs --- and many vertices with only one edge. From our point of view, this kind of network is seen as `favourable' to the propagation of information (systemic risk) and the companies-hubs are systemically relevant.

In conclusion, we can claim that the scale-free character of  the network of relations between the insurers, observed in each period under study, favours the propagation of potential systemic risk in the European insurance sector for which our source is the G-SIIs list of companies. Moreover, due to the shrinkage of the networks during the periods of strong turbulences on financial markets, the ability to propagate significantly increases. 

\section{Acknowledgements}

This article was supported by funds subsided to the Faculty of Finance and Law of the Cracow University of Economics as a support for the research potential.

\end{document}